\documentclass[nohyper,11pt,letterpaper]{JHEP3}

\author{Valeri P. Frolov\\
      Theoretical Physics Institute, Department of Physics\\
      University of Alberta, Edmonton, Canada, T6G 2J1\\
      E-mail: \email{frolov@phys.ualberta.ca}
}

\author{Dmitri V. Fursaev\\
      Bogoliubov Laboratory of Theoretical Physics,
      Joint Institute for Nuclear Research\\
      141980 Dubna, Moscow Region, Russia\\
      E-mail: \email{fursaev@thsun1.jinr.ru}
}

\author{Dejan Stojkovi\'{c}\\
       MCTP, Department of Physics, University of Michigan,\\
       Ann Arbor, MI 48109 USA\\
      E-mail: \email{dejans@umich.edu}
}

\abstract{
We study interaction of rotating higher dimensional
black holes with a brane in  space-times with large extra dimensions.
In the approximation when a black hole is slowly rotating and the
tension of the brane is small we demonstrate that  the black hole
loses some angular momentum to the brane. As a result of this
effect a black hole in its final stationary state can have
only those components of the angular momenta which are connected with
Killing vectors generating transformations preserving a position of
the brane.  The characteristic time when a rotating black hole with
the gravitational radius  $r_0$ reaches this final state
is $T\sim r_0^{p-1}/(G\sigma)$, where $G$ is the higher dimensional
gravitational coupling constant, $\sigma$ is the brane tension, and
$p$ is the number of extra dimensions.
}


\preprint{{\tt gr-qc/0403002}\\
 Alberta-Thy-03-04 ,  \  MCTP-04-11
}

\title{Rotating black holes in brane worlds}

\begin{document}

\section{Introduction}
\setcounter{equation}0

There are several
reasons why brane world scenarios  \cite{ADD,RS}
became popular recently. One of the main reasons is that in these
scenarios the fundamental energy scale can be as low as a few TeV
which opens a possibility of the experimental tests of predictions of
these models in the future collider and cosmic ray experiments
\cite{EG,Land:03}. The most dramatic prediction of these models is a
possibility of creation of mini black holes when the center of mass
energy of two colliding particles becomes higher than a fundamental
energy \cite{EG}.

Being gravitational solitons black holes can propagate in the bulk
space and thus to serve as probes of extra dimensions. We
focus our attention on black holes with a size much smaller than the
size of extra dimensions. In this case the effects  of the
extra dimensions on the black hole geometry are small and can be
neglected. (For recent review of black holes in a space-time with
large extra dimensions see \cite{kanti}.)

In a general case a black hole created by a collision of two
particles is rotating.  If there is no emission of the bulk
gravitons  the bi-plane of the black hole rotation lies within the
brane. In a more realistic situation when  bulk gravitons are
emitted, a black hole can also acquire an angular momentum in the
bi-plane not lying within the brane. There are other processes which
may result in the black hole  rotation in the bulk dimensions. For
example,  if a black hole collides with a particle or another black
hole with emission of bulk gravitons, or when it emits bulk gravitons
in the Hawking evaporation process.

The aim of this paper is to demonstrate that  a rotating black hole
interacting with a brane  loses some of the components of its
angular momenta. We study this effect in the approximation when a
black hole is slowly rotating and the tension of the brane is small.
We demonstrate that as a result of this friction effect a black hole
in its final stationary state can have only those components of the
angular momenta which are connected with Killing vectors generating
transformations preserving a position of the brane. We illustrate
this result first by considering  a 4-dimensional Kerr black hole
interacting with  a thin domain wall. Next we prove this result for a
slowly rotating higher dimensional black holes interacting with
branes.

Our calculations show that the characteristic time of the relaxation
during which a rotating black hole reaches the equilibrium state is
shorter than the time during which it loses its bulk angular momentum
because of the Hawking radiation. This may have important experimental
signature of mini-black holes in future collider and
cosmic ray experiments.

\section{Static black holes}
\setcounter{equation}0

In the approximation when the
gravitational back-reaction of the brane
is neglected its world-sheet obeys the Nambu-Goto equation
\begin{equation}\label{1}
^{(n+1)}\Delta
X^{\mu}+\gamma^{ab}\Gamma^{\mu}_{\ \
\nu\lambda}X^{\nu}_{,a}X^{\lambda}_{,b}=0~~.
\end{equation}
The relations $X^{\mu}=X^{\mu}(\zeta^a)$ determine the embedding of
the $n+1$-dimensional brane into $N+1$-dimensional  bulk space-time.
$\zeta^a$, ($a,b=0,n$), are internal coordinates in the brane and
$X^\mu$, ($\mu=0,N$) are  coordinates in the bulk space with the
metric $g_{\mu\nu}$. The connections $\Gamma^{\mu}_{\ \ \nu\lambda}$
are determined for  $g_{\mu\nu}$.
To exclude degenerate cases we assume that
$0<n<N$. $^{(n+1)}\Delta$ is the box-operator for the
induced metric
\begin{equation}\label{2}
\gamma_{ab}=g_{\mu\nu}\, X^{\mu}_{,a}X^{\nu}_{,b}\, .
\end{equation}
The stress-energy  tensor of the brane is
defined as follows:
\begin{equation}\label{3}
\sqrt{-g}T^{\mu\nu}
=\sigma \int d^{n+1}\zeta \delta^{(N+1)}(X-X(\zeta))\, t^{\mu\nu}
~~\, ,
\end{equation}
where $t^{\mu\nu}=\sqrt{-\gamma} \gamma^{ab}X^{\mu}_{,a}X^{\nu}_{,b}$
and $\sigma$ is the brane's tension.

The metric  of a non-rotating higher dimensional black hole is 
 \cite{Tang}
\begin{equation}\label{4}
d\bar{s}^2=-Bdv^2+2drdv+r^2d\Omega^2_{N-1}\, .
\end{equation}
The coordinate $v$ is the advanced time, $d\Omega^2_{N-1}$ is the
metric on the unit sphere $S^{N-1}$, and $B=1-\left({r_0 /
r}\right)^{N-2}$. For $N=3$ this metric reduces to the Schwarzschild
metric. The gravitational radius $r_0$ is related to the black hole
mass $M$ as follows
\begin{equation}
M={(N-1){\cal A}_{N-1}\over 16\pi G_{N+1}} r_0^{N-2} \, ,
\end{equation}
where ${\cal A}_{N-1}={2\pi^{N/2}\over \Gamma(N/2)}$ is the area of a
unit sphere $S^{N-1}$ and $G_{N+1}$  is the
$N+1$-dimensional gravitational coupling constant which has
dimensionality $[length]^{(N-2)}/[mass]$.

Consider a unit sphere $S^{N-1}$ embedded in a $N$-dimensional
Euclidean space $R^N$, and let $X^A$, ($A=1,\ldots,N$) be the
Cartesian coordinates in $R^N$. One can choose these coordinates so
that the equations $X^{n+1}= \ldots = X^N=0$ determine the
$n$-dimensional hyper-surface (brane). This hyper-surface intersects
the unit sphere $S^{N-1}$ along a surface ${\cal S}$ which has a
geometry of a round unit sphere $S^{n-1}$. The surface ${\cal S}$ is
a higher dimensional analogue of a `large circle' on a
two-dimensional sphere.  In particular, being considered as a
sub-manifold of $S^{N-1}$ it has a vanishing extrinsic curvatures, and
hence is a geodesic sub-manifold. We denote by $\omega^{\alpha}$
coordinates on ${\cal S}$, and by $d\omega_{n-1}^2$ the metric on it.

One can construct a solution for a static $n$-brane as follows. We
use $\zeta^{a}=(v,r,\omega^{\alpha})$ as coordinates on the brane. Then
\begin{equation}\label{7}
d\bar{\gamma}^2=-Bdv^2+2drdv+r^2d\omega^2_{n-1}~~~,
\end{equation}
is the induced geometry on the brane. It
is easy to check that such a surface is geodesic and hence is a
solution of the Nambu-Goto equations (\ref{1}).

Denote by $\xi^{\mu}$ a Killing vector field generating a rotation in
some bi-plane.  Then the flux per unit time $v$ of the corresponding
angular momentum of the  brane, $\dot{J}^{b}$, through the surface
$r$=const is given by the expression
\begin{equation}\label{8a}
\dot{J}^{b}=-\int_{r=\mbox{const}}\sqrt{-g} T^{r\nu}\xi_{\nu}d^{N-1}
\Omega~~.
\end{equation}
Due to the conservation law
$T^{\mu\nu}_{~~~;\nu}=0$ the flux
$\dot{J}^{b}$ does not
depend on $r$.
Let us denote by $\dot{J}$ the rate of the loss of
the angular momentum of the black hole.
The angular momentum is transmitted to the brane
and therefore $\dot{J}=-\dot{J}^{b}$. By using
(\ref{3}), (\ref{8a}) one finds
\begin{equation}\label{8}
\dot{J}=\sigma\int_{r=\mbox{const}}d^{n-1}\omega\,
t^{\mu\nu}n_{\mu} \xi_{\nu}\, ,
\end{equation}
where $n_{\mu}=r_{,\mu}$.
The integral is  over $(n-1)$
dimensional sphere and $d^{n-1}\omega$ is a measure on a unit sphere
$S^{n-1}$.

For a static black hole, since $\xi^{\mu}$ is tangent to the
surface $r$=const,
$\dot{J}=0$.

\section{Kerr black hole}
\setcounter{equation}0

Before considering higher dimensional
rotating black holes let us
discuss a simpler case of a brane attached to the rotating black hole in
the 4-dimensional space-time.
We consider only slowly rotating black holes. For
$a/M\ll 1$ one can write the Kerr metric in the form
\begin{equation}\label{19}
ds^2=d\bar{s}^2-2a\sin^2\theta \, d\varphi \left({r_0 \over r}\,
dv+dr\right)\, .
\end{equation}
Here $d\bar{s}^2$ is the Schwarzschild  metric
\begin{equation}\label{20}
d\bar{s}^2=-Bdv^2+2drdv+r^2(\sin^2\theta d\varphi^2+d\theta^2)\, ,
\end{equation}
and $B=1-r_0/r$, $r_0=2M$.
Denote by $\alpha$ an angle between the axis of rotation and the
brane, then the equation of the unperturbed domain wall is
$\varphi=\bar{\varphi}(\theta)$ where
\begin{equation}\label{21}
\sin\bar{\varphi}=\tan\alpha\, \cot\theta\, ,
\end{equation}
and $\alpha \leq \theta \leq \pi-\alpha$ for $0\leq \alpha \leq
\pi/2$.
The induced metric on the world-sheet of such a tilted domain
wall is
\begin{equation}\label{22}
d\bar{\gamma}^2=-Bdv^2+2drdv+{r^2\sin^2\theta \over \sin^2\theta -
\sin^2\alpha}  d\theta^2\, .
\end{equation}
The domain wall deformed by the black-hole rotation
is described by the equation $\varphi=\bar{\varphi}+\psi$, where
$\psi$ obeys the equation
\begin{equation}\label{23}
^{(3)}\bar{\Delta }\psi+
{2 \over r^2}\cot\theta \psi_{,\theta}+2{B \over r}\psi_{,r}= {a \over
 r^3}\, ,
\end{equation}
where $^{(3)}\bar{\Delta }$ is the box-operator in the metric
(\ref{22}). This equation has a solution $\psi=-a/r$. It is possible to
show that this is a unique solution which is regular both at the
horizon and infinity \cite{FFS}.

For this regular
solution, $T^r_{\, \varphi}$ is given by the following expression
\begin{equation}\label{24}
\sqrt{-g} T^{r}_{\, \varphi}=-
\sigma r_0\,a \sin\theta\sqrt{\sin^2\theta-\sin^2\alpha}
\delta(\varphi-\bar{\varphi}(\theta))
\, .
\end{equation}
This quantity is already of the first order in $a$ and one can
use (\ref{8a}) to obtain
\begin{equation}\label{25}
\dot{J}=-\pi \sigma a r_0\cos^2\alpha=-2\pi G_4 \sigma \cos^2\alpha J\, .
\end{equation}
The angular momentum flux vanishes when the domain wall is in the
equatorial plane of the rotating black hole \cite{1}. Thus this is the
final stationary equilibrium
configuration of the rotating black hole in the presence of the
domain wall. The relaxation time when the black hole reaches this
final state is $T\sim (\pi G_4 \sigma \cos
^2\alpha)^{-1}$.

For a cosmic string attached to the Kerr black hole, a similar
problem can be solved for an arbitrary value of the rotation
parameter $a$ because  solutions of the Nambu-Goto equations are
known  exactly \cite{FrHeLa}.  In this case the final stationary
configuration is a string directed along the rotation axis. The
relaxation time is $T\sim r_0/(4G_4 \sigma  \sin^2\alpha)$, where
$\alpha$ is an initial angle between the the string and the axis of
rotation \cite{FFS}.

\section{Higher dimensional rotating black holes}
\setcounter{equation}0

Now we consider
a general case. We assume that a $N$-dimensional
rotating black hole is attached to a $n$-dimensional brane. If the
black hole size is much smaller that the size of extra dimensions,
and the tension of the brane is small, the gravitational field of the
black hole is described by the Myers-Perry (MP) metric
\cite{MP}, \cite{2}. This metric beside the time-like at infinity Killing
vector $\xi_{(t)}^{\mu}$ has $[N/2]$ (the integer part of $N/2$)
mutually commutative and mutually orthogonal  Killing vectors
$\xi_{(i)}^{\mu}$ singled out by the property that they have closed
integral lines.
The Killing vectors $\xi_{(i)}$ are elements of the Cartan
sub-algebra of the group of rotations $SO(N)$.
The MP metric is characterized by the gravitational
radius $r_+$ and by $[N/2]$  rotation parameters $a_i$. Such a black
hole has angular velocities $\Omega_i=a_i/(r_+^2+a_i^2)$. The vector
$\eta=\xi_{(t)}+\sum_i \Omega_i \xi_{(i)}$  on the horizon
becomes a null generator of the horizon.
(The summation over $i$ is performed from $i=1$ to
$i=[N/2]$.)

For slow rotation $a_i/r_0\ll 1$, the MP metric in the Kerr-incoming
coordinates takes the form
\begin{equation}\label{27}
ds^2=d\bar{s}^2- {2\over r^2}[dr+
(r_0/r)^{N-2}dv]\, \varrho_{\mu}dx^{\mu}\, ,
\end{equation}
\begin{equation}\label{varrho}
\varrho^{\mu}=\sum_i a_i \xi_{(i)}^{\mu}\, ,
\end{equation}
where $d\bar{s}^2$ is the unperturbed metric
(\ref{4}). For this form of the metric relations ${\cal
L}_{\xi_{(i)}}\bar{g}_{\mu\nu}=0$ imply that ${\cal
L}_{\xi_{(i)}}g_{\mu\nu}=0$.
The angular  momenta of the black hole  $J_i$ are defined as
\begin{equation}\label{28}
J_i={{\cal A}_{N-1}\over 8\pi G_{N+1}} r_0^{N-2}\, a_i={2\over N-1} Ma_i\,
 .
\end{equation}
In the linear approximation (\ref{27}) $r_+\approx r_0$,
$\Omega_i\approx a_i/r_0^2$, and $\eta =\xi_{(t)}+r_0^{-2}\varrho$.

Consider a static brane in the metric $d\bar{s}^2$. We analyze now
what happens with the brane in the presence of slow rotation.
Interaction of the brane with the black hole results in the change of
the brane position $X^{\mu}=\bar{X}^{\mu}+\delta X^{\mu}$. By
linearizing the Nambu-Goto equations (\ref{1}) one obtains a linear
equation for $\delta X^{\mu}$. It is possible to show that a
solution of these equations which is regular at the horizon and
infinity is
\cite{FFS}
\begin{equation}\label{perx}
\delta X^{\mu}=\psi(r)\, \varrho^{\mu}\, ,\hspace{0.3cm}
\psi'(r)=(1-(r_0/r)^{n-1})/(r^2B)\, .
\end{equation}
We calculate now the rate of the loss of the
angular momentum of the black hole which interacts
with a stationary brane. Since for the static metric
$d\bar{s}^2$ $\dot{J}_i$
vanishes, it is sufficient to calculate the variation of (\ref{8})
induced by the metric perturbations. We have
\begin{equation}\label{29}
\dot{J}_i=\sigma\int_{r=\mbox{const}}d^{n-1}\omega\,
\delta(t^{\mu\nu}n_{\mu}\xi_{(i)\nu})\, .
\end{equation}
It is easy to check that in the linear in $a_i$ approximation the
following relations for the variations induced by the perturbed metric
(\ref{27}) are valid: $\delta(\sqrt{-\gamma})=0$,
\begin{equation}\label{a}
\xi_{(i) a}\delta\gamma^{ra}= {1\over r^2}\,
(\xi^{\parallel}_{(i)},\varrho^{\parallel})-B(\xi^{\parallel}_{(i)},
\delta X_{,r}^{\parallel})\, ,
\end{equation}
\begin{equation}
\gamma^{ra}\delta\xi_{(i)a}=- {a_i\over r^2}\,  (\xi_{(i)})^2\, ,
\end{equation}
\begin{equation}\label{b}
\gamma^{ra}\, \xi_{(i) \lambda} \delta X^{\lambda}_{,a}=B
(\xi_{(i)},\delta X_{,r})\, .
\end{equation}
We denote by $p^{\parallel}$ a projection of the vector $p$ on the
brane.
$(p,q)$ is a scalar product of vectors $p$ and $q$ in the
unperturbed metric, $(p,q)=\bar{g}_{\mu\nu}p^{\mu}q^{\nu}$.

The flux of the angular momentum from the black hole to the brane
changes the angular momenta of the black hole (\ref{28}). In the
linear approximation the equations for the
change of the angular momenta of the black hole can be written as
follows
\begin{equation}\label{dyn}
\dot{{\bf J}}=-{\bf K}\, {\bf J}\, .
\end{equation}
We use bold-faced quantities for vectors and tensors in the space of
rotation parameters, so that  ${\bf J}$ and  ${\bf K}$ have
components  $J_i$,  and $K_{ij}$.  Relations (\ref{28})--(\ref{b})
enable one to get ${\bf K}$  in the form
\begin{equation}\label{K}
K_{ij}=(N-1)\sigma r_0^{n-1}k_{ij}/(2M)\, ,
\end{equation}
\begin{equation}\label{int}
k_{ij}=\int_{S^{n-1}} d\omega^{n-1}\,
{1\over r^2}\, (\xi^{\perp}_{(i)},\xi^{\perp}_{(j)})\, .
\end{equation}
We denote by $p^{\perp}$  a projection of a vector $p$  orthogonal to
the brane. In an agreement with the conservation law, the angular
momentum flux does not depend on the radius $r$ of the surface where
it is calculated.

Note that in the linear in $a_i$ approximation
$\dot{M}=\dot{r}_0=0$ and $M$ and $r_0$ in (\ref{K}) are considered
as constant parameters. The evolution equation (\ref{28}) can also be
written as $\dot{{\bf a}} =-{\bf K}\, {\bf a}$, where $\bf a$ is a
vector with components $a_i$.
This equation shows
that  the black hole can
be stationary, $\dot{\bf a}=0$, if and only if ${\bf a}$ is the zero
vector, ${\bf K a}=0$. In this case the equation ${\bf a}^{T}{\bf K
a}=0$ implies that
\begin{equation}
\int_{S^{n-1}} d\omega^{n-1}\, (\varrho^{\perp},\varrho^{\perp})=0\, ,
\end{equation}
and hence $\varrho^{\perp}=0$. This means that the corresponding
Killing vector $\varrho$ generates transformations that preserve a
position of the brane. The stationary metric of the final black hole
configuration in this case is given by (\ref{27}) where $\varrho$ is
a  vector tangent to the brane.  Because in the considered
approximation $\varrho$ is related to
the null generator of the black hole horizon,
$\eta=\xi_{(t)}+r_0^{-2}\varrho$, one can also describe the final state
of the black hole as a state where $\eta$ is tangent to the
brane world-sheet.

From (\ref{perx}) it follows that in the final stationary state  $\delta
X^{\perp \mu}=0$. Since the tangent to the brane components of
$\delta X^{\mu}$ can always be gauged away, the brane in this case
is not deformed.

Beside zero  eigenvectors, the non-trivial matrix ${\bf K}$ has
eigenvectors with positive eigenvalues. These eigenvectors define the
directions in the space of parameters $a_i$ for which the evolution
is  damping. The  damping is caused by the `friction' which is a
result of the interaction between the black hole and the brane. The
characteristic time of the relaxation process during which the black
hole reaches its final state is
\begin{equation}
T\sim r_0^{p-1}/ (G_{N+1}\sigma)\, \sim T_* ({r_0/ L_*})^{p-1}\,
(\sigma_*/\sigma)\, .
\end{equation}
Here $p=N-n$ is the number of extra dimensions,
$\sigma_*=M_*/L_*^n$ and quantities
$M_*$, $L_*$ are, respectively,
the fundamental mass and length of the theory.

The black hole can also lose its bulk components of the rotation by
emitting  Hawking quanta in the bulk.  The characteristic time of this
process is $T_{H}\sim T_* (r_0/L_*)^N$. For black holes which can be
treated classically  $r_0\gg L_*$, $T_H \gg T$. Thus the friction
effect induced by the brane is the dominant one.

\section{Discussion}
\setcounter{equation}0

We considered interaction of rotating black  holes
with  branes. Such systems include several physically interesting
examples, such as cosmic strings and thin domain walls interacting
with the Kerr black hole, as well as rotating black holes in a brane
world.  In the slow rotation approximation we demonstrated that
there exist an angular momentum transfer from a black hole to the
attached brane. It vanishes when the generator of rotation
$\varrho$, (\ref{varrho}), is tangent to the brane. One can expect
that a similar result may be valid beyond the adopted approximation,
that is when a black hole is not slowly rotating and  the brane
generates a nontrivial gravitational field. It will happen for
example if  a higher dimensional analogue of the Hawking theorem
\cite{Hawk:72} is valid.

We would like to conclude the paper by the following remark. We
focused our attention on the higher dimensional space-times   with
vanishing bulk cosmological constant (ADD model \cite{ADD}).  A
similar problem concerning general properties of higher dimensional
rotating black holes with the horizon radius $r_{+}$ can be
addressed  in the Randall-Sundram (RS) models \cite{RS} provided  the
bulk  cosmological constant is much smaller than $r_{+}^{-2}$.  A
characteristic property of such models is the existence of $Z_2$
symmetry. Under $Z_2$ transformation the brane remains unchanged,
while the components of any vector orthogonal to the brane change
their sign.  Thus $Z_2$ symmetry implies $\varrho^{\perp}=0$.  Hence
a stationary black hole attached to the brane in the RS-model can
rotate only within the brane. Exact solutions describing rotating
black holes on two-branes \cite{EHM} possess this property.

The relaxation process related to the presence of $\varrho^{\perp}$
which is typical for the ADD-model is absent in the
RS-model. This is an additional signature which in principle may
allow one to distinguish between these models in observations.

\bigskip
\vspace{12pt} {\bf Acknowledgements}:\ \

V.F. and D.F. kindly acknowledge the support from  the NATO
Collaborative Linkage Grant (979723). The work of V.F. and D.F. is
also partially supported   by the Killam Trust and the Natural
Sciences and Engineering Research Council of Canada. DS thanks the DOE
and the Michigan Center for Theoretical Physics for support at the
University of Michigan.


\begin{thebibliography}{99}
\bibitem{1} It is easy to
check that for an arbitrary value of $a$ the equatorial plane of the
Kerr metric is a geodesic surface and hence it obeys the
Nambu-Goto equations. For this configuration  there is no angular
momentum transfer.
\bibitem{2} In  the higher dimensional case there is no uniqueness
theorem and there may exist rotating black hole solutions different
from the MP metric. An example is a black ring solution
\cite{EmRe:02}. This metric has no slow rotation limit. For this
reason we do not discuss it in the paper. Also in the slow rotation
limit the effect of instability of rapidly rotating black holes
\cite{EmMy:03} is not important.
\end{thebibliography}
\end{document}